\documentclass[twocolumn]{aastex63}
\usepackage{amsmath}
\shorttitle{An alternative channel to BH LMXBs: dynamical friction of dark matter? }
\shortauthors{Qin \& Chen}

\begin{document}


\title{An Alternative Channel to Black Hole Low-Mass X-ray Binaries: Dynamical Friction of Dark Matter?  }

\author[0000-0001-9206-1641]{Ke Qin}
 \affil{School of Science, Qingdao University of Technology, Qingdao 266525, People's Republic of China; chenwc@pku.edu.cn}
\affil{School of Physics, Zhengzhou University, Zhengzhou 450001, People's Republic of China}
\author[0000-0002-0785-5349]{Wen-Cong Chen}
  \affil{School of Science, Qingdao University of Technology, Qingdao 266525, People's Republic of China; chenwc@pku.edu.cn}
  \affil{School of Physics and Electrical Information, Shangqiu Normal University, Shangqiu 476000, People's Republic of China}



\begin{abstract}
Both the anomalous magnetic braking of Ap/Bp stars and the surrounding circumbinary disk models can account for the formation of black hole (BH) low-mass X-ray binaries (LMXBs), while the simulated effective temperatures of the donor stars are significantly higher than the observed values. Therefore, the formation of BH LMXBs is not still completely understood. In this work, we diagnose whether the dynamical friction between dark matter and the companion stars can drive BH binaries to evolve toward the observed BH LMXBs and alleviate the effective temperature problem. Assuming that there exists a density spike of dark matter around BH, the dynamical friction can produce an efficient angular momentum loss, driving BH binaries with an intermediate-mass companion star to evolve into BH LMXBs for a spike index higher than $\gamma = 1.58$. Our detailed stellar evolution models show that the calculated effective temperatures can match the observed value of most BH LMXBs for a spike index range of $\gamma = 1.7-2.1$. However, the simulated mass-transfer rates when $\gamma = 2.0$ and $2.1$ are too high to be consistent with the observed properties that BH LMXBs appears as soft X-ray transients. Therefore, the dynamical friction of dark matter can only alleviate the effective temperature problem of those BH LMXBs with a relatively short orbital period.
\end{abstract}


\section{Introduction}
Black holes (BHs) are compact objects with gravitational masses exceeding the maximum value $\sim2-3~M_{\odot}$ of a neutron star \citep{rhoa74},  which are evolutionary products of massive stars with masses of $\ga 20~M_{\odot}$ \citep{frye12}. To date, it discovered about two dozen BHs in the Milky Way, all of them were found in X-ray binaries where their dynamical masses can be availably constrained \citep{remi06,casa14,shao20}. Most BHs locate BH X-ray binaries with low-mass (usually $< 1.0~M_{\odot}$) donor stars and short orbital periods (usually $< 1.0$ days), which are generally named as BH low-mass X-ray binaries \citep[LMXBs;][]{lee02,pods03,ritt03}. Due to the thermal-viscous instability of the accretion disk, all BH LMXBs appear as transients in observation. Taking into account a duty cycle, the total number of BH LMXBs in the Milky Way is estimated to be in the range of $100 - 1000$ \citep{tana95,mccl06}. BH LMXBs are ideal probes for testing the evolution of massive stars, binary stars, and common envelopes (CE).

At present, the formation channels of BH LMXBs mainly include the triple star channel and the classical isolated binary evolution channel. To account for the formation of BH LMXB A0620-00, \cite{eggl86} proposed a triple star scenario, in which a BH (or neutron star) can be formed in a close massive binary that orbits a distant late dwarf in a triple system. Subsequently, two spiral-in phases may result in the formation of a BH LMXB with a short period. \cite{naoz16} investigated the formation of BH LMXBs
via three-body gravitational dynamics, and found three distinct evolutionary channels for BH LMXB candidates including "eccentric", "giant", and "classical" channels by a large Monte-Carlo sample of simulations. In their simulated candidates, 8\% systems evolved from the "classical" channel, and eventually form BH LMXBs with short orbital periods. Recently, the canonical BH LMXB V404 Cygni is confirmed to be an inner binary with a tertiary companion at a separation greater than 3500 AU \citep{burd24}, providing a robust evidence of the triple star channel.

In the isolated binary evolution channel, it is generally believed that BH LMXBs evolved from those primordial binaries consisting of a massive star and an intermediate/low-mass companion \citep{li15}. To evolve to a close orbit, it is inevitable to experience a CE evolutionary phase \citep{pacz76,deko87}. However, it is still controversial whether the low-mass companion star can provide sufficient orbital energy to eject the envelope of the massive BH progenitor during the CE phase \citep{port97,kalo99,pods03} \footnote{\cite{wang16} demonstrated that BH LMXBs could evolve from those primordial binaries with a low-mass companion via the standard CE channel if most BHs are born through failed supernovae.}. Some works invoked some potential additional sources of energy to eject the envelope during the CE phase \citep{pods10,ivan11,ivan15}. If the primordial binary contains an intermediate-mass secondary star, the orbital energy problem of the CE phase can be successfully solved. The observation of the BH LMXB XTE J1118+480 hinted the signature of CNO-processed elements \citep{hasw02}, indicating the progenitor of the donor star should be an intermediate-mass star. Nevertheless, the orbital period should increase when the mass is transferred from the less massive donor star to the more massive BH because magnetic braking was generally thought to be absent in the intermediate-mass star without a convective envelope \citep{kawa88}. The existence of BH LMXBs implies that there is indeed an additional orbital angular-momentum-loss (AML) mechanism during the evolution of BH intermediate-mass X-ray binaries (IMXBs).

About 5 \% intermediate-mass stars (so-called Ap/Bp stars) possess anomalously strong magnetic fields of $10^{2}-10^{4}~\rm G$, \cite{just06} proposed the coupling between the magnetic field and the donor-star winds induced by X-ray irradiation can produce anomalous magnetic braking, driving the BH IMXBs to evolve into BH LMXBs. As an alternative mechanism, the tidal torque between a circumbinary (CB) disk and the binary system can also efficiently extract the orbital angular momentum from BH IMXBs, driving them to evolve toward BH LMXBs \citep{chen06,chen15}. Especially, the CB disk model can successfully account for the observed orbital period derivatives in three BH LMXBs \citep{chen15,chen19}. The binary evolution models based on the CB disk or anomalous magnetic braking models can perfectly explain the compact orbits of BH LMXBs, however, the simulated effective temperatures of donor stars are significantly higher than the observed values \citep{just06,chen06,chen19}. Therefore, the formation of BH LMXBs is not still completely understood. It also implies that some unknown physical mechanisms dominate the formation and evolution of BH LMXBs.

\section{Dynamical Friction Model of Dark Matter}
Cosmological simulations found that dark matter distributes in galactic halos according to the Navarro-Frenk-White profile \citep{nava96,bert10}. If a supermassive BH at the center of a galaxy grows purely adiabatically by the standard accretion
of dust and gases, it could alter the dark matter profile and produce a high-density cusp of dark matter, i.e. the dark matter spike \citep{gond99,gned04,merr04,sade13}. Furthermore, dark matter minispikes may more easily form around a spinning intermediate-massive BH with a mass of $10^{3-5}~M_{\odot}$ \citep{ferr17}. In general, the density profile of these dark matter spikes abides by a simple power-law spherically symmetric distribution as $\rho_{\rm DM}\propto r^{-\gamma}$, where $r$ is the distance from the center BH, $\gamma$ is the spike index \citep{gond99,eda13,kava20}. In principle, the spike index can be derived according to the model of adiabatic growth of supermassive BHs \citep{youn80}. As a consequence, the spike index ranges from 1.5 to 2.5, depending on the model \citep{gned04,merr04,sade13,fiel14,lacr18}.

A dark matter density spike would result in a dynamical friction effect, which could influence the extreme-mass-ratio inspiral processes. When a star is orbiting inside a collisionless dark matter background, it produces a gravitational force to exert on the dark matter particles. Conversely, the cluster of the dark matter particles locating behind the star exerts an inverse gravitational force on the star, which would slow down the star, resulting in an effect of so-called dynamical friction \citep{chan43}. The influence of dynamical friction on the intermediate- or extreme-mass-ratio inspirals around a supermassive BH \citep{anto12,li22,spee22} or an intermediate-mass BH \citep{eda15,yue19,kava20,beck22,dai22} are investigated extensively. As a member of BH population, is it possible whether stellar mass BHs can also create dark matter spikes like supermassive or intermediate-mass BHs? Recently, the dynamical friction due to a dark matter density spike was proposed to be responsible for the anomalous fast orbital decays in two BH LMXBs XTE J1118+480 and A0620-00 \citep{chan23}. Nevertheless, they only presented an analytical estimation for the effect of dynamical friction on the orbital evolution of these two BH LMXBs in the current stage. Heretofore, a detailed stellar evolution modeling on the BH binaries including the dynamical friction of dark matter has been missing in the literature. In this work, we perform such a simulation for the first time.

When the donor star in a BH LMXB orbits in the dark matter background, the dynamical friction exerted in the donor star dissipates its orbital energy. According to the gravitational force due to dynamical friction, the orbital energy dissipation rate can be written as \citep{yue19,chan23}
\begin{equation}
\dot{E}_{\rm DF}=-\frac{4\pi G^{2} \mu^{2} \rho_{\rm DM}\xi(\sigma)\ln \Lambda}{v},
\end{equation}
where $G$ is the gravitational constant, $\mu$ is the reduced mass of the binary, and $\xi(\sigma)$ is a numerical factor relating to the distribution function and the velocity dispersion $\sigma$ of dark matter, ${\rm ln} \Lambda \approx {\rm ln} (\sqrt{M_{\rm BH}/M_{\rm d}})$ is the Coulomb logarithm \citep[$M_{\rm BH}$ and $M_{\rm d}$ are the masses of the BH and the donor star,][]{kava20}, $v$ is the orbital velocity of the donor star. Similar to \cite{chan23}, we take $\xi(\sigma)= 1$.

The density of dark matter around a BH is a piecewise function as follows \citep{lacr18}
\begin{equation}
\rho_{\rm DM}=\left\{
\begin{array}{lll}
0 & {\rm when} & r\le 2R_{\rm s}\\
\rho_{\rm 0}(r/r_{\rm sp})^{-\gamma} & {\rm when} & 2R_{\rm s}<r\le r_{\rm sp}\\
\rho_{\rm 0} & {\rm when} & r> r_{\rm sp}\\
\end{array}\right.
\end{equation}
where $R_{\rm s}=2GM_{\rm BH}/c^{2}$ is the Schwarzschild radius of the BH, and $r_{\rm sp}$ is the spike radius of dark matter. Similar to \cite{fiel14} and \cite{eda15}, we adopt a standard assumption $r_{\rm sp} = 0.2 r_{\rm in}$, where $r_{\rm in}$ is the influence radius of the BH. When the radial distance is greater than $r_{\rm sp}$, the dark matter density becomes a constant distribution with a density of $\rho_{\rm 0}$, which is related to the position of the system in the Milky Way. The distribution of dark matter in the Milky Way follows the Navarro-Frenk-White dark matter density profile \citep{nava96}. Except for the high-density region at the Galactic center, the discrepancy of the dark matter density is negligible in the other region of the Galactic disk \citep{mcmi17}. For simplicity, we take a uniform local dark matter density as $\rho_{0}\approx\rho_{\odot}= 0.33\pm 0.03~\rm GeV\,cm^{-3}$, where $\rho_{\odot}$ is the dark-matter density at the solar position \citep{abli20}.

Same to \cite{merr03} and \cite{merr04}, the influence radius of the BH is calculated by
\begin{equation}
M_{\rm DM}(r\le r_{\rm in})=\int_{0}^{r_{\rm in}} 4\pi r^{2}\rho_{\rm DM}{\rm d}r = 2M_{\rm BH}.
\end{equation}
Consequently, the spike radius $r_{\rm sp}$ is a function of the BH mass and continuously changes during the Roche lobe overflow (RLOF). It is impossible to derive an analytical solution of $r_{\rm in}$ (or $r_{\rm sp}$). Therefore, we obtain a numerical solution of $r_{\rm in}$ by an iterative numerical calculation.

The rate of orbital AML caused by the dynamical friction can be expressed as follows
\begin{equation}
\dot{J}_{\rm DF}=\frac{\dot{E}}{\Omega}= -\frac{G^{2} \mu^{2}(1+q)\rho_{\rm DM}\xi(\sigma)\ln \Lambda P_{\rm orb}^{2}}{\pi a},
\end{equation}
where $\Omega$, $a$, $q=M_{\rm d}/M_{\rm BH}$, and $P_{\rm orb}$ are the angular velocity, orbital separation, mass ratio, and orbital period of the binary, respectively. The drag force received by the donor star due to dynamical friction is $f_{\rm DF}=4\pi G^{2} \mu^{2} \rho_{\rm DM}\xi(\sigma)\ln \Lambda/v^{2}$ \citep{yue19}, and the force arm of the drag force is $l_{\rm DF}=a/(1+q)$ respect to the mass center of the binary. Therefore, the spin-down torque produced by the dynamical friction is $T_{\rm DF}=f_{\rm DF}l_{\rm DF}$, which would derive the same rate of orbital AML as Equation (4).

\section{Stellar Evolution Code}
The progenitors of BH LMXBs are thought to be binary systems consisting of a stellar mass BH (with an initial mass $M_{\rm BH,i}$) and an intermediate-mass companion star (with an initial mass $M_{\rm d,i}$) on a circular orbit (with an initial orbital period $P_{\rm orb,i}$). We use the binary module in the Modules for Experiments in Stellar Astrophysics \citep[MESA,version r12115,][]{paxt11,paxt13,paxt15,paxt18,paxt19} to model the evolution of BH binaries. The initial system contains a stellar-mass BH and an intermediate-mass companion star with a solar composition (i.e., $X = 0.7, Y = 0.28, Z = 0.02$). The code only models the nuclear synthesis and evolution of the intermediate-mass companion star, and the BH is considered a point mass. When the companion star fills its Roche lobes, the mass transfer is calculated by adopting the "Kolb" mass-transfer scheme \citep{kolb90}. The code will iteratively simulate the evolution of BH binaries until the stellar age is greater than the Hubble timescale (14 Gyr) or the time step is less than the default minimum time step limit.

During the evolution, the total orbital AML of BH binaries is given by
\begin{equation}
\dot{J}=\dot{J}_{\rm DF}+\dot{J}_{\rm GR}+\dot{J}_{\rm MB}+\dot{J}_{\rm ML},
\end{equation}
where $\dot{J}_{\rm DF},\dot{J}_{\rm GR},\dot{J}_{\rm MB}$, and $\dot{J}_{\rm ML}$ are the rates of orbital AML induced by the dynamical friction, gravitational wave (GW) radiation, magnetic braking, and mass loss, respectively. Among these mechanisms, both $\dot{J}_{\rm DF}$ and $\dot{J}_{\rm GR}$ always operate during the evolution of BH binaries. However, the magnetic braking mechanism only takes effect when the donor star possesses both a convective envelope and a radiative core, and we adopt a standard magnetic braking scenario \citep{rapp83} and a magnetic braking index $\gamma = 4$ \citep{verb81}. Moreover, the mass growth of the BH is limited by the Eddington accretion rate, and the excess material is thought to carry away the specific orbital angular momentum of the BH. Our inlists are available on Zenodo: 10.5281/zenodo.11218828.

\section{Results}

To show the influence of dynamical friction on the evolution of BH binaries, we plot the rate of orbital AML due to dynamical friction and its fraction in the total rate of orbital AML as a function of the orbital period and the donor-star mass for a BH binary with $M_{\rm BH,i} = 10~M_{\odot}$, $M_{\rm d,i} = 3~M_{\odot}$, $P_{\rm orb,i} = 3.0$ days and $\gamma = 1.6$ and 2.0 in Figure 1. A high spike index naturally leads to a large dark-matter density and a large $\dot{J}_{\rm DF}$. At a same orbital period or donor-star mass, $\dot{J}_{\rm DF}$ with $\gamma = 2.0$ is approximately four orders of magnitude larger than that with $\gamma = 1.6$. Because of an extremely large $\dot{J}_{\rm DF}$ ($\sim10^{38}-10^{40}~\rm g\,cm^{2}s^{-2}$), dynamical friction with $\gamma = 2.0$ contributes a fraction of $\ga95\%$ in the total rate of orbital AML. According to Equation (4), $\dot{J}_{\rm DF}\propto P^{4/3}$. With the orbital shrinkage, the rate of orbital AML due to dynamical friction slightly decreases before the mass transfer starts (this phenomenon is invisible for $\gamma = 2.0$ because a short timescale of pre-mass transfer). Once the donor star fills its Roche lobe, a high mass-transfer rate dominates the orbital evolution of the BH X-ray binary. The orbital period slowly increases because the mass is transferred from the less massive donor star to the more massive BH. In the early stage of the mass transfer, $\mu$ and $q$ decrease, while $a$, $P$, and $\rho_{\rm DM}$ increase. Our calculations indicate that $\dot{J}_{\rm DF}$ continuously decreases, which implies that $\dot{J}_{\rm DF}$ is very sensitive to the evolution of $\mu$ and $q$ according to Equation (4). According to Equation (2), the density of dark matter is greater at a smaller radii, while the rate of orbital AML eventually appears as a decreasing tendency during the orbital decay. This also implies that the mass change of the donor star, resulting in a change of $\mu$ and $q$, plays a vital role in influencing the dynamical friction. As $\gamma=1.6$, the detailed changes in the total evolution are as follows: the $\dot{J}_{\rm DF}$ caused by dynamical friction is reduced by 5 orders of magnitude, the mass term ($\mu^{2}(1+q)\ln \Lambda$) alters by 5 orders of magnitude, the dark-matter density ($\rho_{\rm DM}$) increases by only 2 orders of magnitude,  and the spike radius ($r_{\rm sp}$) varies by less than one-tenth.

\begin{figure}
\centering
\includegraphics[width=1.0\linewidth,trim={0 0 0 0},clip]{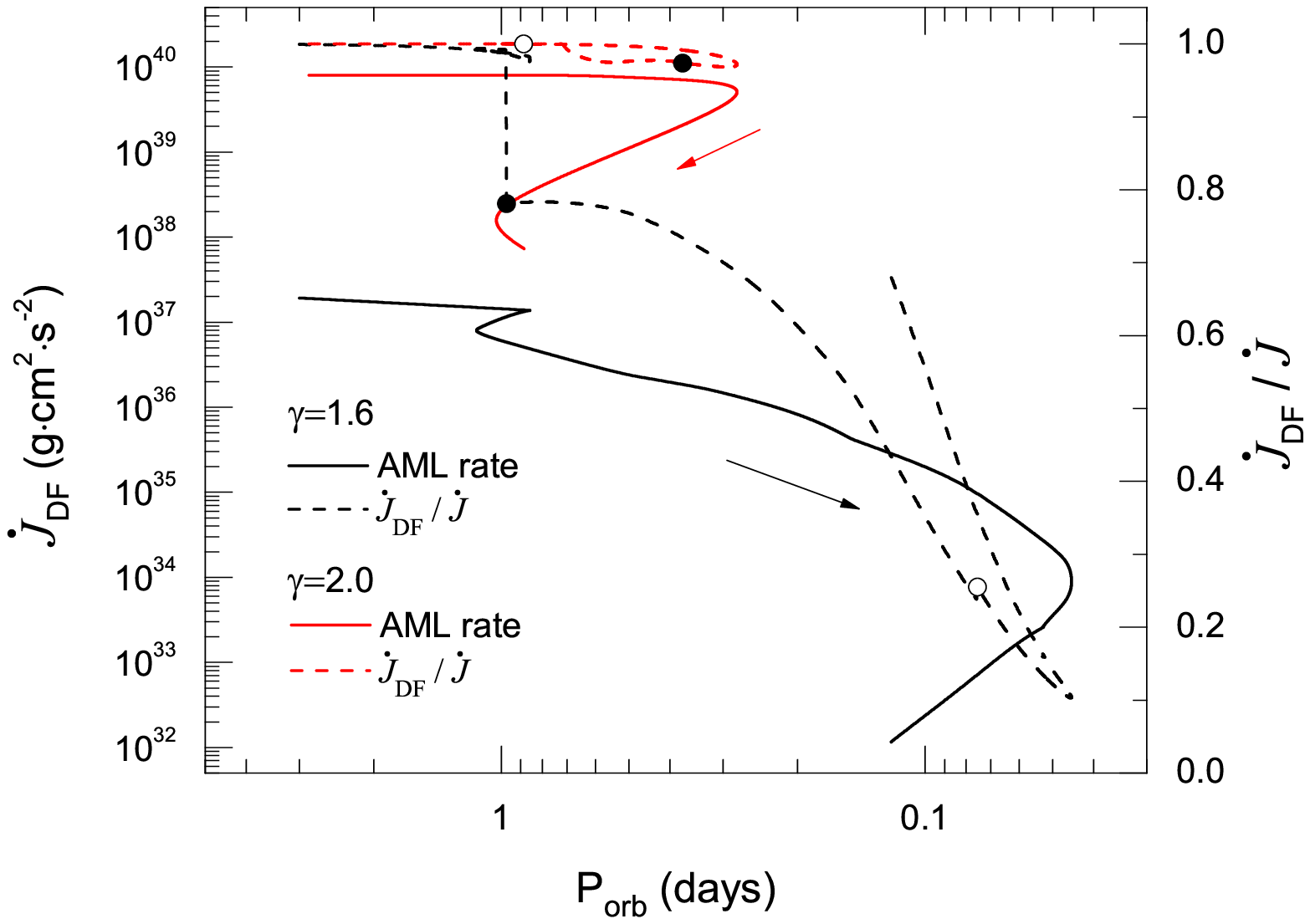}
\includegraphics[width=1.0\linewidth,trim={0 0 0 0},clip]{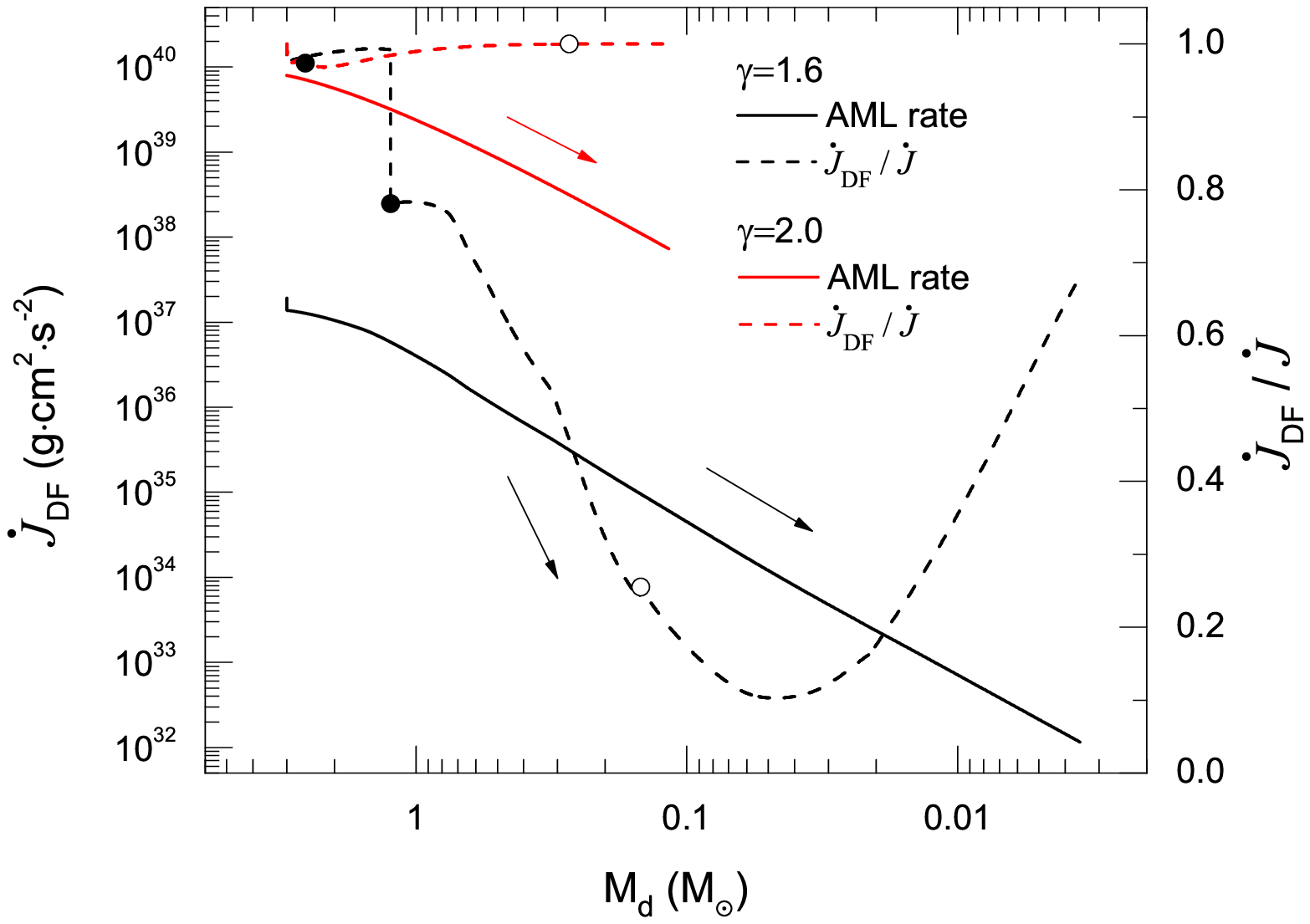}
\caption{Evolution of a BH binary with $M_{\rm BH,i} = 10~M_{\odot}$, $M_{\rm d,i} = 3~M_{\odot}$, $P_{\rm orb,i} = 3.0$ days and $\gamma = 1.6$ and $2.0$ in the rate of orbital AML (or the fraction) due to dynamical friction vs. orbital period diagram (top panel) and the rate orbital AML (or the fraction) due to dynamical friction vs. the donor-star mass diagram (bottom panel). The solid line represents the rate of orbital AML caused by dynamical friction, and the dashed line represents its fraction in the total rate of orbital AML. The arrows correspond to the evolutionary direction. The solid and open circles mark the onset and end of magnetic braking, respectively.} \label{fig:orbmass}
\end{figure}

It is noteworthy that the influence of dynamical friction on the orbital evolution is efficient for a long orbital period rather than a short orbital period. When $\gamma=1.6$, the dynamical friction dominates the orbital evolution of a BH LMXB with an orbital period longer than 0.2 days, providing a fraction of $\ga 60\%$ in the total rate of orbital AML. When the donor star develops a convective envelope at a mass of $\sim1.5~M_{\odot}$, magnetic braking contributes a fraction of $\sim20\%$ in the total rate of orbital AML. The subsequent orbital evolution is dominated by both dynamical friction and magnetic braking, and the fraction provided by dynamical friction in the total rate of orbital AML gradually declines. As the cut-off of magnetic braking, the fraction contributing by dynamical friction decreases to $\sim20\%$. Subsequently, the GW radiation drives the BH X-ray binary to evolve into a BH ultracompact X-ray binary. At the minimum orbital period of $0.045~\rm days$, the fraction contributing by dynamical friction reaches to a minimum fraction of $\sim0.1$. With the increase of the orbital period, the fraction contributing by GW radiation continuously reduces, and the dynamical friction dominates the orbital evolution of the BH X-ray binary again. The rates of orbital AML via both GW radiation and dynamical friction are related to the orbital separation. However, an increasing orbital separation produces a decreasing $\dot{J}_{\rm GR}$ and a increasing $\dot{J}_{\rm DF}$ because $\dot{J}_{\rm GR}\propto a^{-7/2}$ and $\dot{J}_{\rm DF}\propto a^{2}$, respectively.

\begin{figure}
\centering
\includegraphics[width=1.15\linewidth,trim={0 0 0 0},clip]{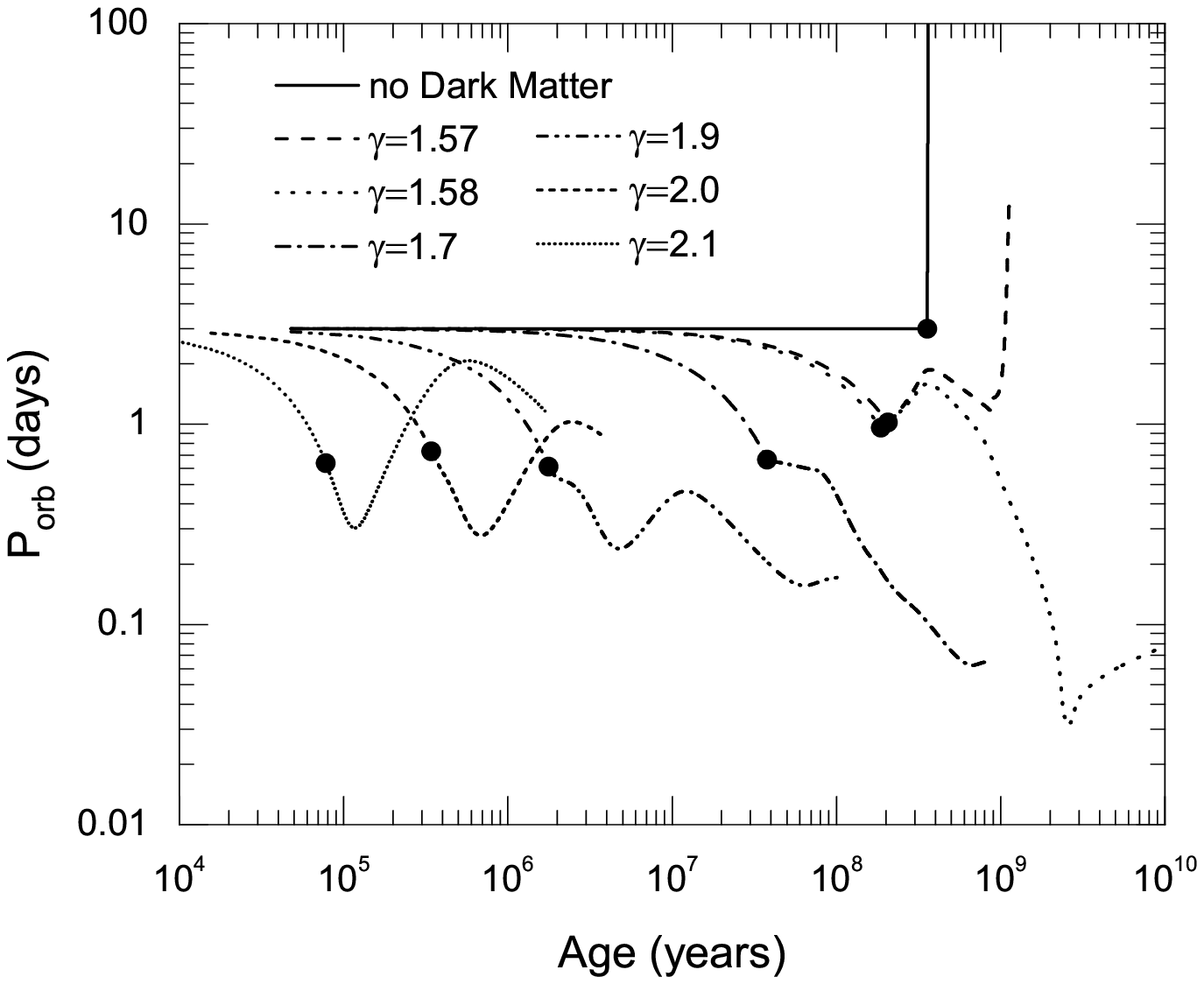}
\includegraphics[width=1.15\linewidth,trim={0 0 0 0},clip]{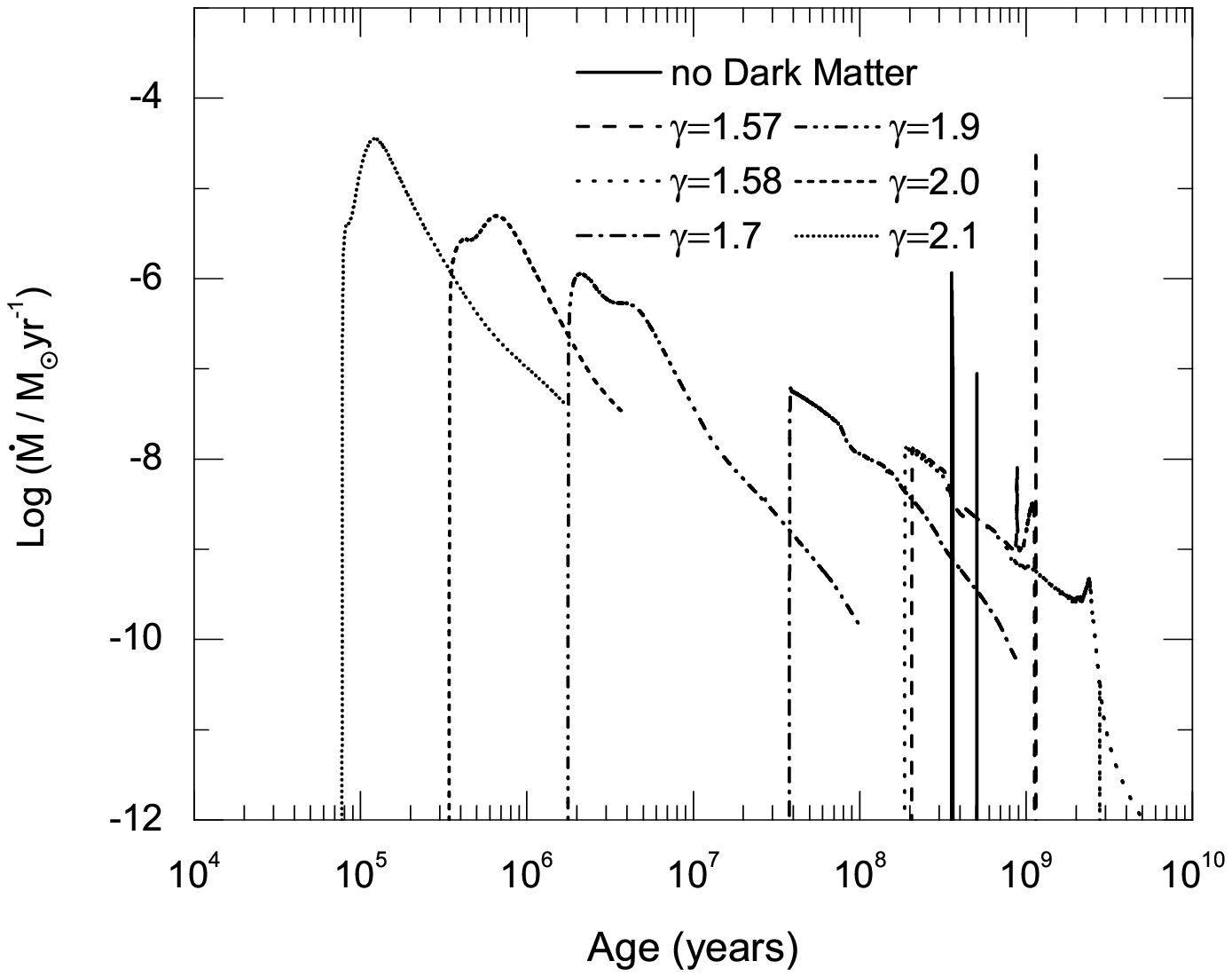}
\caption{Evolution of a BH binary with $M_{\rm BH,i} = 10~M_{\odot}$, $M_{\rm d,i} = 3~M_{\odot}$, and $P_{\rm orb,i} = 3.0$ days in the orbital period vs. stellar age diagram (top panel) and the mass-transfer rate vs. stellar age diagram (bottom panel). The solid, dashed, dotted, dashed-dotted, dashed-dotted-dotted, short-dashed, and short-dotted curves indicate the evolutionary tracks of the cases without dynamical friction, and with $\gamma=1.57$, 1.58, 1.7, 1.9, 2.0, and 2.1, respectively. The solid circles represent the onset of mass transfer.} \label{fig:orbmass}
\end{figure}

Figure 2 depicts the evolution of orbital periods and mass-transfer rates with the stellar age for a BH binary with $M_{\rm BH,i} = 10~M_{\odot}$ , $M_{\rm d,i} = 3~M_{\odot}$, $P_{\rm orb,i} = 3.0$ days under different spike indices and the case without dynamical friction. Because of the absence of magnetic braking, the orbital period of the BH binary without the dynamical friction is almost constant before the RLOF occurs. With the beginning of the mass transfer, the orbit continuously expands because the mass is transferred from the less massive donor star to the more massive BH. Therefore, it is impossible to form those BH LMXBs if they lack an efficient loss mechanism of AML like anomalous magnetic braking \citep{just06} or a surrounding CB disk \citep{chen06}. According to Equation (2), a high spike index $\gamma$ naturally produces a high density of dark matter at the position of the donor star. Under the same parameters of the BH X-ray binary, the dynamical friction with a high spike index would drive an efficient orbital AML, giving rise to a rapid orbital shrinkage and an early mass-transfer phase. Our models find that there exists a critical spike index $\gamma_{\rm cr}=1.58$, over which the BH binary could evolve into a BH LMXB.

As $\gamma =1.7, 1.9, 2.0$, and 2.1, those BH binaries initiate mass transfer respectively at the stellar age of $\sim40, 1.8, 0.33$, and $0.08$ Myr, while the beginning age of mass transfer is $186.4$ Myr when $\gamma = 1.58$. A high spike index $\gamma$ leads to a high mass-transfer rate, and a short lifetime in BH X-ray binary stage. Especially, those BH X-ray binaries with $\gamma=1.9, 2.0$, and $2.1$ produce a high mass-transfer rate of $\ga 10^{-7}~M_{\odot}\,\rm yr^{-1}$ in a timescale of $\sim1-5~\rm Myr$, appearing as ultraluminous X-ray sources without high-mass donor stars \citep[$\ga10~M_{\odot}$, see also][]{rapp05}. BH LMXBs with a short orbital period generally appear as soft X-ray transients in observations \citep{king96,king97a,king97b}. However, our simulated mass-transfer rate with $\gamma = 2.0$ and $2.1$ are extremely high, and are highly incompatible with their transient properties. The lifetimes of five BH X-ray binaries with $\gamma=1.58,1.7, 1.9, 2.0$, and 2.1 are $\sim5000, 1000, 100, 3$, and 1 Myr respectively, which have a difference of approximately four orders of magnitude. In the case of $\gamma = 1.58$, the orbital shrinkage due to dynamical friction cannot compensate for the rapid orbital expansion caused by a high mass-transfer rate, hence the orbital period firstly increases until the magnetic braking starts. Among those spike indices not more than $\gamma_{\rm cr}$, a spike index that is very close to $\gamma_{\rm cr}$ tends to form a BH LMXB with a relatively short minimum orbital period. This phenomenon is because the BH binary with a small spike index would evolve a long timescale, leading to a high He abundance in the donor-star core, which naturally forms a more compact donor star and a correspondingly shorter orbital period \citep{tutu87,lin11}. Considering the dynamical friction, those BH binaries with an intermediate-mass donor star and a spike index of $\gamma=1.58-1.7$ can evolve toward BH ultracompact X-ray binaries \citep[$P_{\rm orb} < 1.5~\rm hours$;][]{vanh13}, however, this evolutionary tendency can not achieve for the standard magnetic braking model \citep{qin23}. A spike index higher than 1.9 is impossible to produce BH ultracompact X-ray binaries.

\begin{figure}
\centering
\includegraphics[width=1.15\linewidth,trim={0 0 0 0},clip]{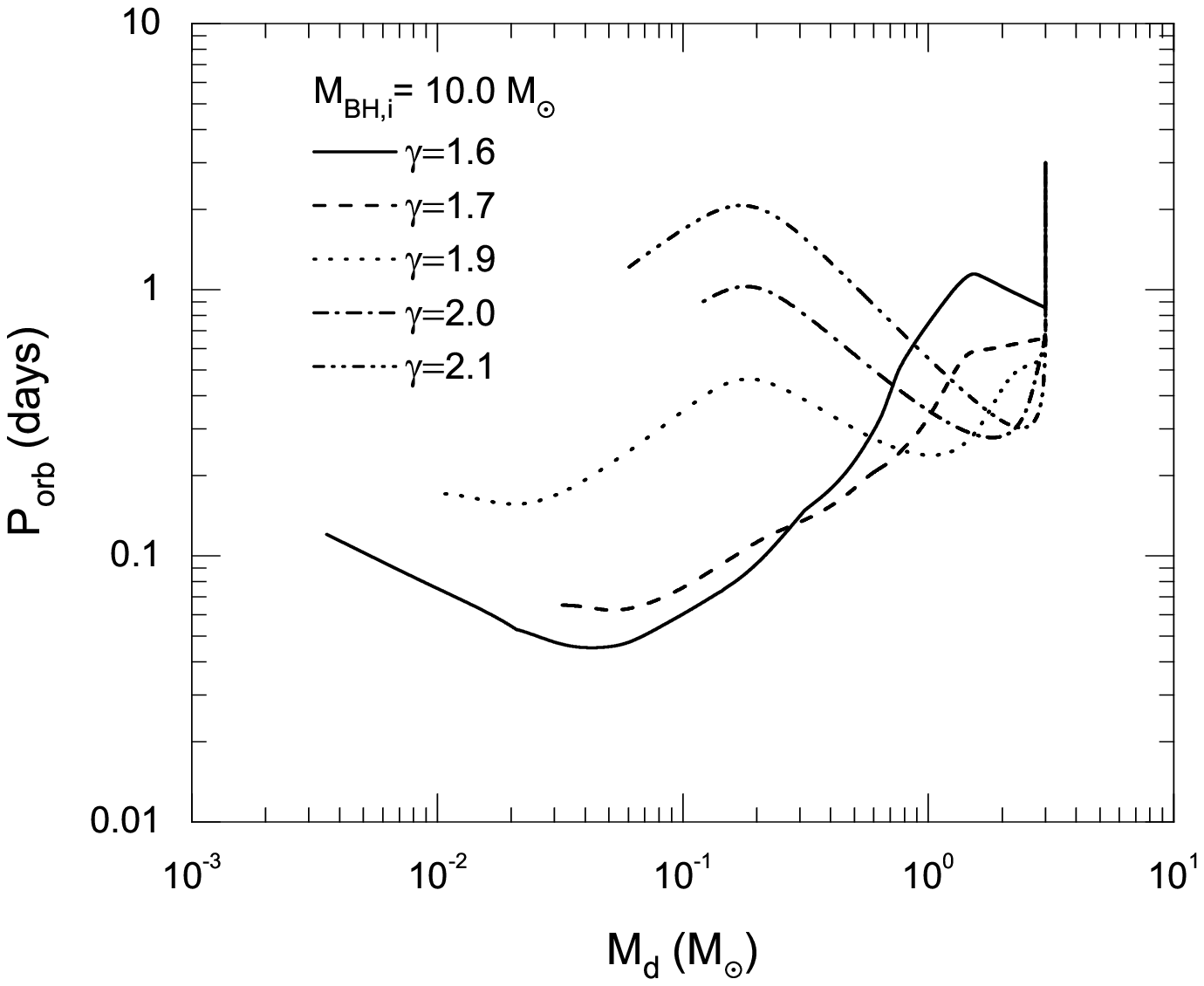}
\includegraphics[width=1.15\linewidth,trim={0 0 0 0},clip]{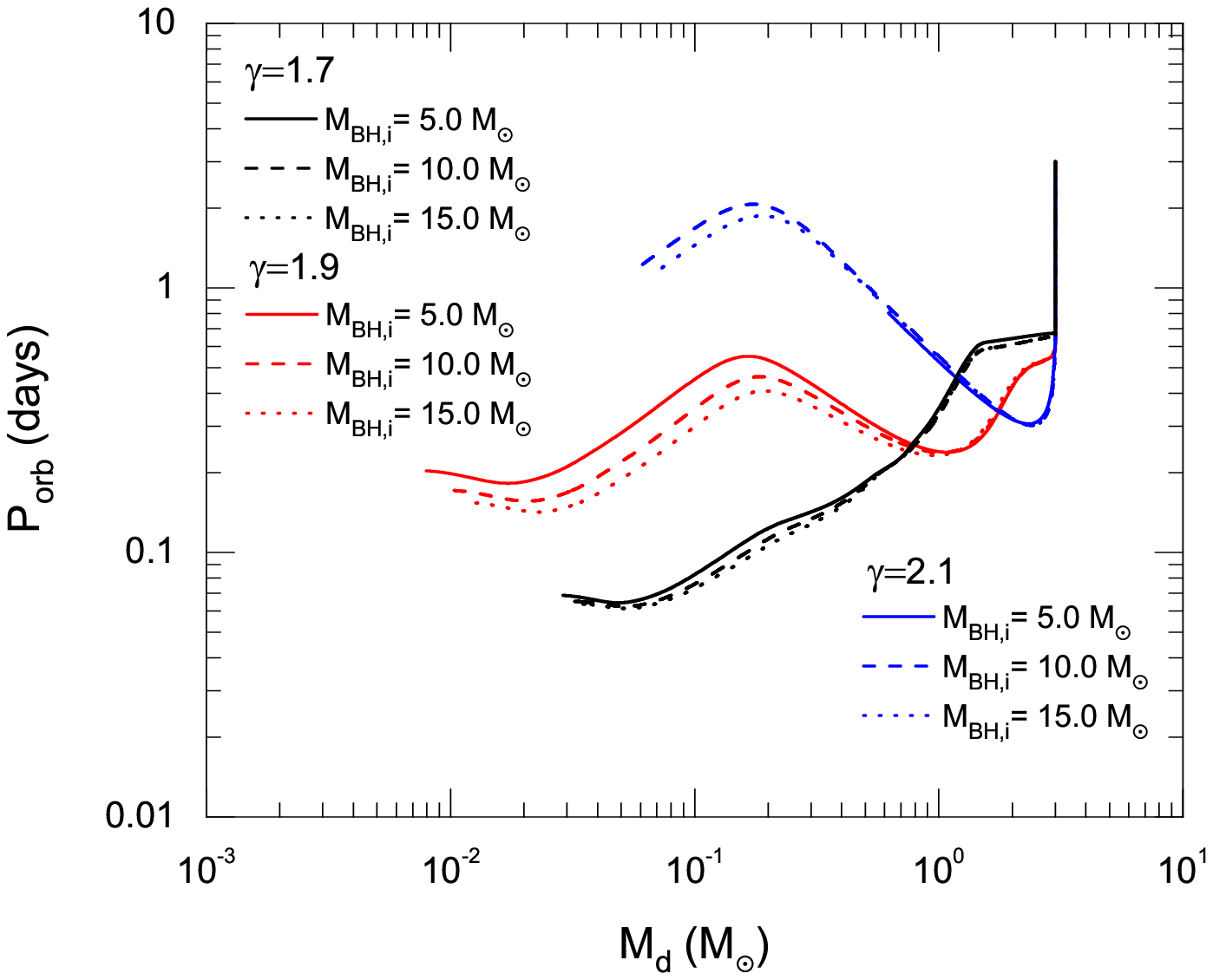}
\caption{Evolution of BH binaries with $M_{\rm d,i} = 3~ M_{\odot}$, $P_{\rm orb,i} = 3.0$ days under different $M_{\rm BH,i}$ and $\gamma$ in the orbital period vs. the mass of donor star diagram.} \label{fig:orbmass}
\end{figure}

Figure 3 describes the evolution of BH binaries with $M_{\rm d,i} = 3~ M_{\odot}$, $P_{\rm orb,i} = 3.0$ days under different $M_{\rm BH,i}$ and $\gamma$ in the orbital period versus the donor-star mass diagram. According to Equation (3), the spike radius ($r_{\rm sp}$) relies on the BH mass. As a result, both the BH mass and the spike index $\gamma$ determine the density distribution of dark matter and also influence the rate of orbital AML and the evolution of X-ray binary. Same to Figure 2, a small $\gamma$ tends to produce a short minimum orbital period. The influence of the initial BH mass is not significant on the evolution of BH X-ray binaries. According to Equation (3), a high initial BH mass results in a large influence radius and spike radius of dark matter. For the same orbital separation, $\rho_{0}$, and spike index, it can derive a high dark-matter density at the position of the donor star from Equation (2) and a high rate of orbital AML from Equation (4). As a consequence, a high initial BH mass tends to form a BH LMXB with a relatively short orbital period for the same donor-star mass.

\begin{figure}
\centering
\includegraphics[width=1.15\linewidth,trim={0 0 0 0},clip]{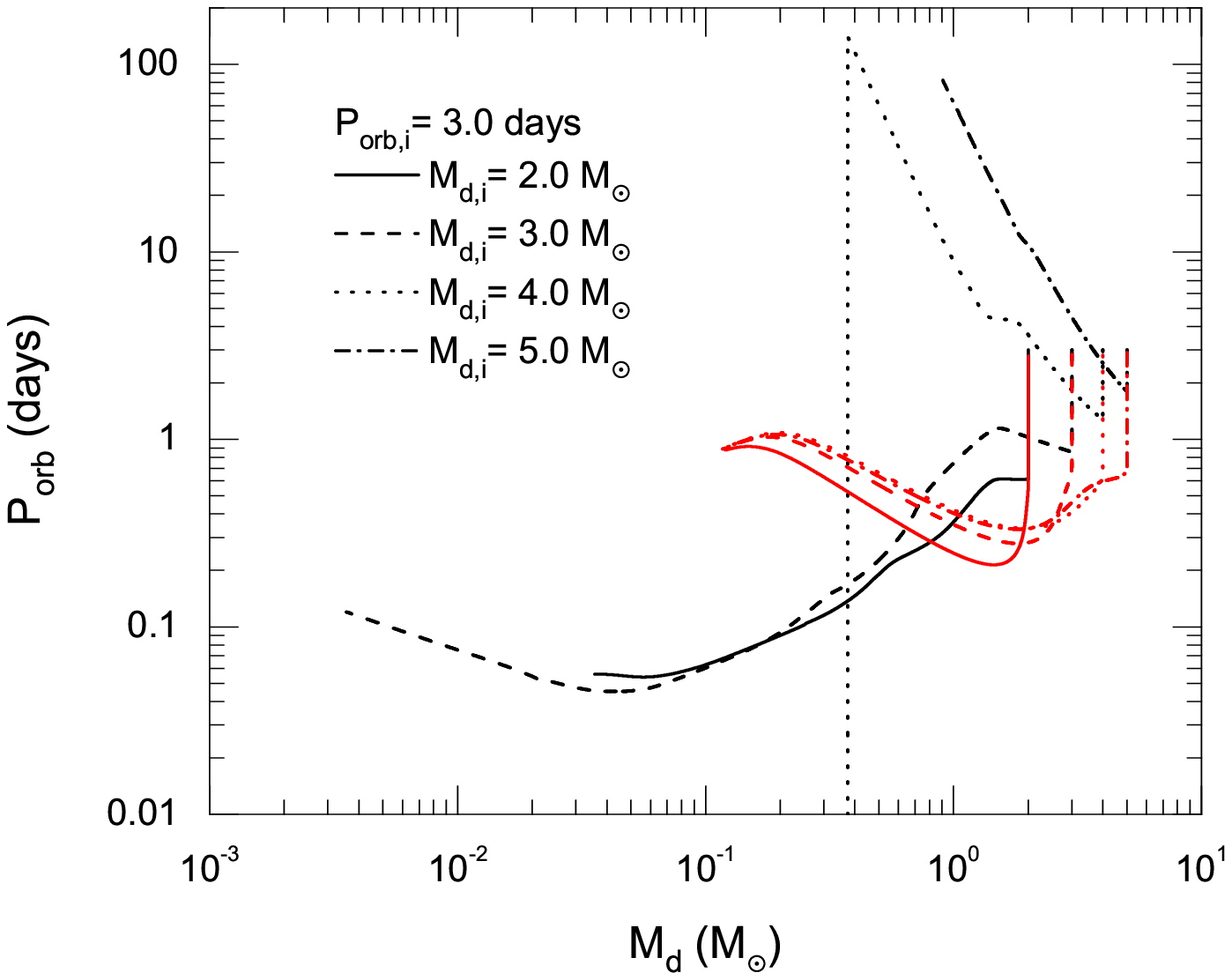}
\includegraphics[width=1.15\linewidth,trim={0 0 0 0},clip]{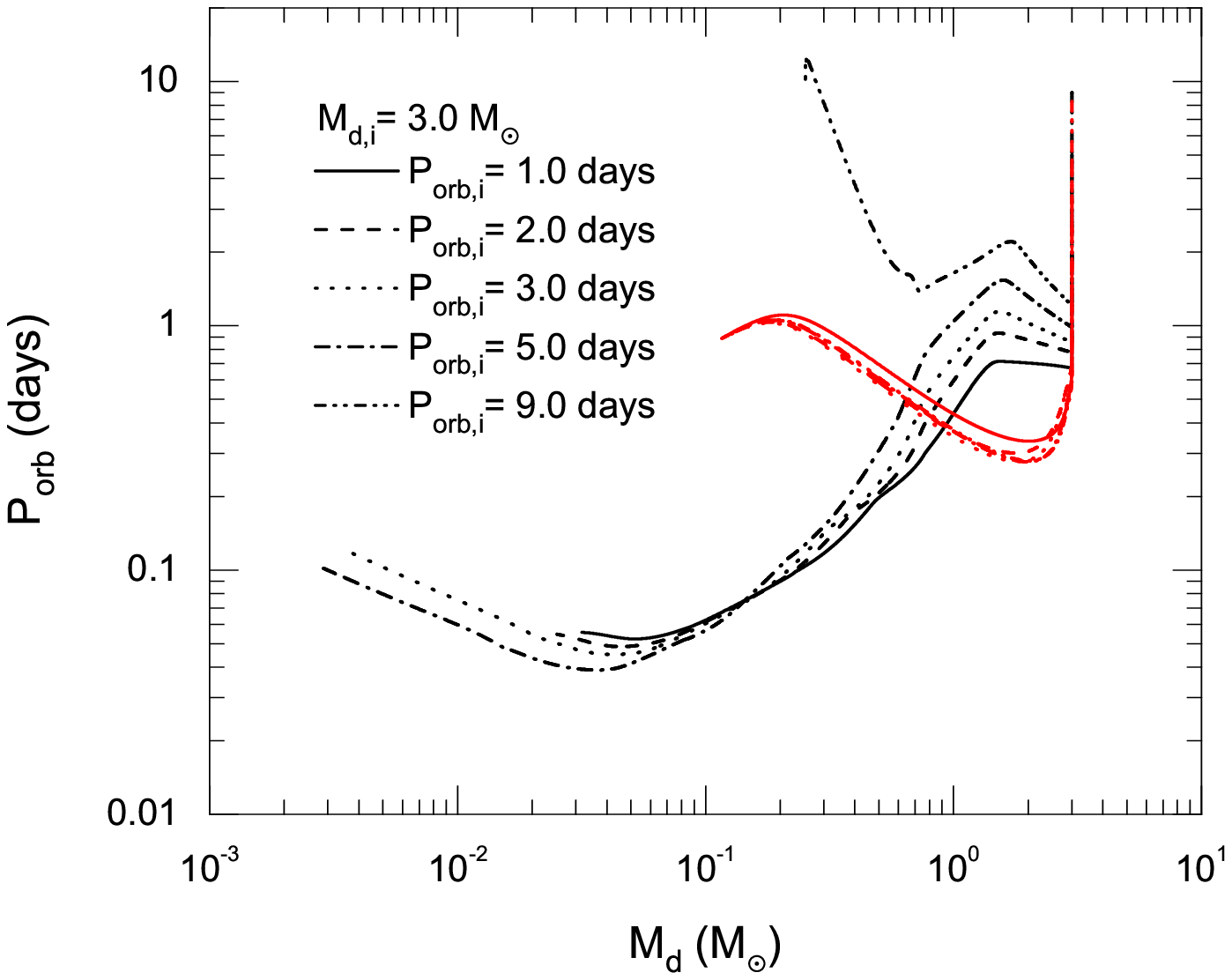}
\caption{Evolution of BH binaries with $M_{\rm BH,i} = 10~M_{\odot}$ and $\gamma = 1.6$ (black curve group) and $2.0$ (red curve group) in the orbital period vs. the mass of donor star diagram under different $M_{\rm d,i}$ (upper panel) and $P_{\rm orb,i}$ (bottom panel). } \label{fig:orbmass}
\end{figure}

Figure 4 summarizes the influence of initial donor-star masses and initial orbital periods on the evolution of BH X-ray binaries when $M_{\rm BH,i} = 10 M_{\odot}$ and $\gamma = 1.6$ and $2.0$. When $\gamma = 1.6$, two BH binaries with an initial companion star of $2.0~ M_{\odot}$ and $3.0~ M_{\odot}$ can evolve into BH LMXBs, while BH binaries with heavy companion stars would evolve toward long-period BH X-ray binaries. This phenomenon arises from the competition between the mass transfer and the AML due to the dynamical friction, in which the former causes the orbit to widen, while the latter gives rise to a shrinking orbit. A heavy donor star provides a high mass-transfer rate and produces a high positive period derivative, which can overcome the negative period derivative due to dynamical friction. Therefore, the orbits of those BH X-ray binaries with heavy donor stars appear as an expansion tendency. As $\gamma = 1.6$, those BH binaries with $M_{\rm BH,i} = 10~M_{\odot}$, $M_{\rm d,i}=3.0~M_{\odot}$, and $P_{\rm orb,i}=1.0-5.0$ days can evolve into BH LMXBs. A long initial orbital period would achieve a short minimum orbital period (the reason is similar to the influence of $\gamma$ on the minimum orbital period). However, the dynamical friction cannot drive a BH binary with an initial period longer than 9.0 days to become detectable BH LMXBs. Therefore, there also exists a so-called bifurcation period forming BH LMXBs, over which BH binaries cannot evolve toward short-period systems \citep{qin23}. As $\gamma = 2.0$, all BH binaries with different initial donor-star masses and initial orbital periods can evolve toward BH LMXBs. Meanwhile, the influence of the initial orbital period on the evolution of BH X-ray binaries with $\gamma = 2.0$ is trivial because of an efficient AML.

\begin{figure}
\centering
\includegraphics[width=1.15\linewidth,trim={0 0 0 0},clip]{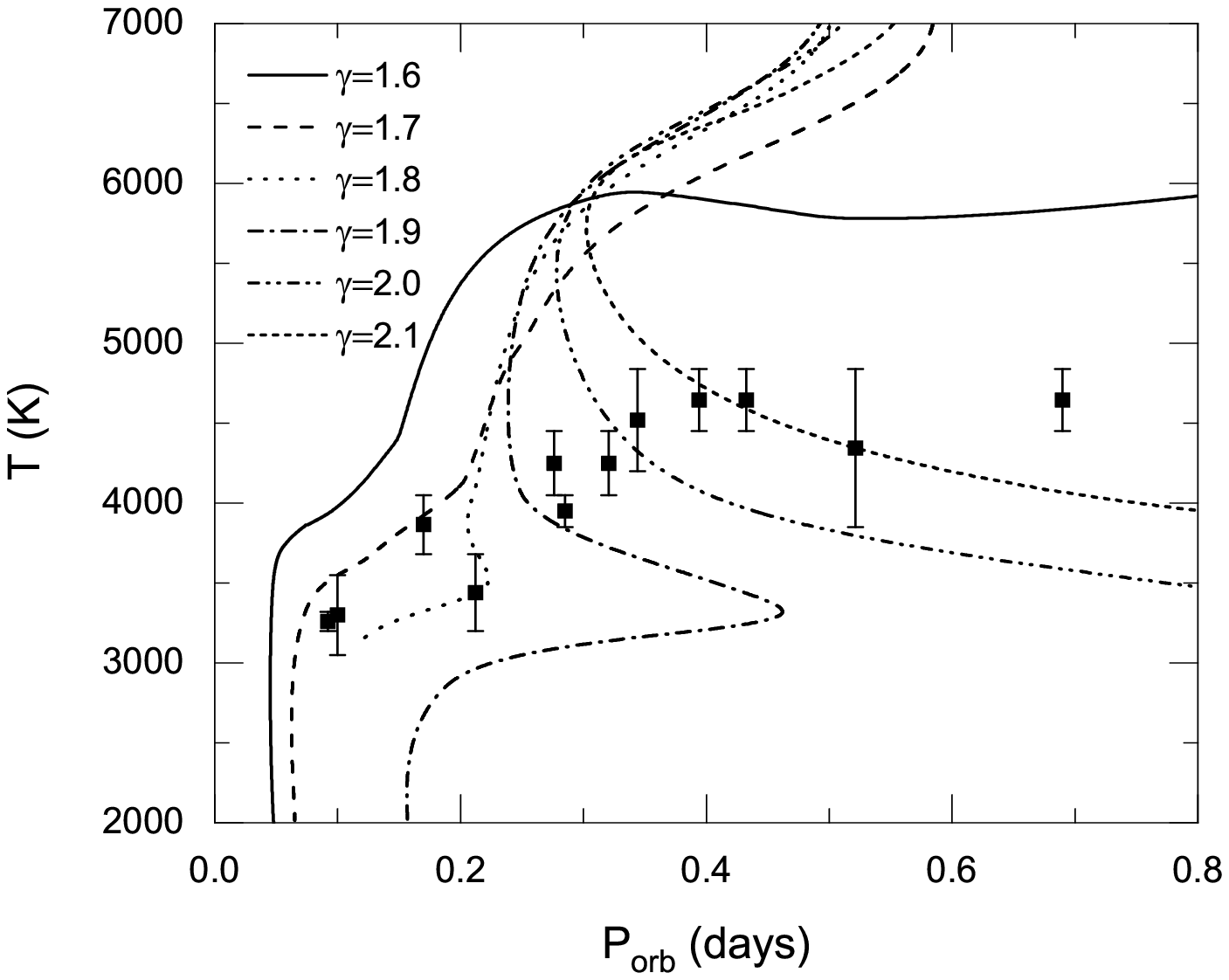}
\caption{Evolution of BH binaries with $M_{\rm BH,i} = 10~M_{\odot}$, $M_{\rm d,i} = 3~M_{\odot}$, $P_{\rm orb,i} = 3.0$ days, and different spike indices $\gamma$ in the effective temperature of the donor star vs. orbital period diagram. The squares with error bars represent the observed BH LMXBs (see also Table 1).} \label{fig:orbmass}
\end{figure}
Table 1 lists the orbital periods, spectral types, and inferred effective temperatures of twelve Galactic BH LMXBs. To test the dynamical friction model, in Figure 5 we illustrate the evolutionary tracks of BH binaries with different spike indices $\gamma$ in the effective temperature of the donor star versus orbital period diagram. For those spike indices in the range from 1.7 to 2.1, our models can match the observed effective temperatures of the donor stars of eight BH LMXBs at the current orbital periods (the other three sources except for MAXI J1820+070 can also be achieved by altering the spike index). When $\gamma \le 1.6$, the simulated effective temperatures are significantly higher than the observed values of all BH LMXBs. In our simulation, XTE J1118+480 is in the period-decreasing stage with $\gamma=1.7$, while A0620-00 and GRS 1124-68 are in the period-increasing stage with $\gamma\sim2.0$ and $\gamma=2.1$, respectively. However, these three sources were detected a rapid orbital decay \citep{gonz14,gonz17}. If the orbital decay is a long-term phenomenon, the dynamical friction can not be responsible for the formation of A0620-00 and GRS 1124-68. To account for the detected orbital period derivatives, the required spike indices of XTE J1118+480 and A0620-00 are $\gamma=1.85^{+0.04}_{-0.04}$ and $\gamma=1.71^{+0.01}_{-0.02}$ \citep{chan23}. Therefore, our detailed stellar evolution models are not consistent with the prediction given by \cite{chan23}. Our simulations find that BH LMXBs with long orbital periods ($>0.3~\rm days$) tend to require a high spike index $\gamma=2.0$ or $2.1$. A small spike index could not only produce short orbital periods but also result in the formation of donor stars with low effective temperatures. Furthermore, the calculated mass-transfer rates when $\gamma = 2.0$ and $2.1$ are highly incompatible with the observed properties that BH LMXBs generally appears as soft X-ray transients. Therefore, the dynamical friction of dark matter can only alleviate the effective temperature problem of the donor stars in those BH LMXBs with a short orbital period $<0.3~\rm days$.

\begin{table*}
\begin{center}
\caption{Observed Parameters of Twelve Galactic BH LMXBs \label{tbl-2}}
\begin{tabular}{@{}lllll@{}}
\hline\hline\noalign{\smallskip}
Source & $P_{\rm orb}$ (days) & Spectral Type & $T_{\rm eff}$ (K) & References \\
\hline\noalign{\smallskip}
MAXI J1820+070 & 0.69 & K3V-K5V & 4450-4840 & \cite{torr20}\\
H1705-250 & 0.52 & K3V-M0V & 3850-4840 & \cite{harl97}\\
GRS 1124-68 & 0.43 & K3V-K5V & 4450-4840 & \cite{geli01}\\
MAXI J1305-704 & 0.39 & K3V-K5V & 4450-4840 & \cite{mata21}\\
GS 2000+251 & 0.34 & K3V-K6V & 4200-4840 & \cite{harl96}\\
A0620-00 & 0.32 & K5V-K7V & 4050-4450 & \cite{cant10}\\
GRS 1009-45 & 0.29 & K7V-M0V & 3850-4050 & \cite{fili99}\\
XTE J1859+226 & 0.28 & K5V-K7V & 4050-4450 & \cite{yane22}\\
GRO J0422+32 & 0.21 & M1V-M4V & 3200-3680 & \cite{harl99}\\
XTE J1118+480 & 0.17 & K7V-M1V & 3680-4050 & \cite{khar13}\\
MAXI J1659-352 & 0.10 & M2V-M5V & 3050-3550 & \cite{kuul13}\\
MAXI J0637-430 & 0.09 & M3.4V-M4V & 3200-3320 & \cite{sori22}\\
\hline\noalign{\smallskip}
\end{tabular}
\tablenotetext{}{}{Note. The columns list (in order): the source name, orbital period, companion's surface effective temperature which is inferred from the given spectral type, references.}
\end{center}
\end{table*}

\section{Discussion}
\subsection{Detection of dark-matter spike}
If dark matter particles are weakly interacting massive particles (WIMPs), WIMP annihilation or decay can produce gamma rays that may be detected by the Fermi Large Area Telescope \citep{acke13}. Especially, WIMP annihilations inside the spike are enhanced and produce a detectable strong gamma-ray photon flux \citep{eda15}. The accurate measurements of cosmic ray electron flux by the Dark Matter Particle Explorer (DAMPE) discovered a sharp peak near 1.4 TeV \citep{ambr17}. \cite{chan19} proposed that the enhanced dark matter annihilation via the $e^{+}e^{-}$ channel around the closest BH A0620-00 can yield a plausible amount of electron and positron flux to interpret the DAMPE 1.4 TeV peak.

The gravitational effects of the dark-matter spike could contribute some hints to the GW signals produced by intermediate-mass-ratio inspirals (IMRI). The environmental effects including dark-matter spikes
could alter the GW signals because those systems can be detected by the space GW detectors like LISA in a mission of several years \citep{mukh23}. For example, the gravitational potential of the dark-matter minispike around intermediate-massive BHs can modify the GW waveform of an IMRI detected by LISA even if dark matter does not annihilate, resulting in a deviation between the observed waveform and a standard template \citep{eda13}. However, it is worth emphasizing that the GW observation can only diagnose the dark matter effect when its profile is steep enough ($\gamma\geq2.0$) \citep{eda13}. Considering the dynamical friction of the dark matter, the dark-matter parameters can be very accurately determined from the GW waveform of an IMRI \citep{eda15}. A smaller mass of the stellar mass object, that of the IMBH, or the larger spike index tends to produce an accurate measurement of dark matter parameters. Even a moderately flatter spike index of 1.7 can still determine the spike index to 10 \% accuracy. Compared the effect of accretion on the GW waveform with those of gravitational pulling and dynamical friction, \cite{yue18} demonstrated that dynamical friction is the dominant mechanism.

BH ultracompact X-ray binaries evolving from the main-sequence star and the He star channels are potential Galactic low-frequency GW sources \citep{qin23,qin24}, which could be detected by LISA \citep{amar23}, TianQin \citep{luo16}, and Taiji \citep{ruan20}.
Meanwhile, the accurate observations of optical and X-ray observations in BH LMXBs can measure the most important physical
parameters, thus, BH LMXBs are the candidate probes for unveiling the dark matter density spike. Our simulations indicate that those BH binaries could evolve into BH ultracompact X-ray binaries when $\gamma=1.7$, which is the lower limit that the GW detection can determine the spike index \citep{eda15}. Therefore, it is possible that the space-borne GW detectors could detect the waveforms from some BH ultracompact X-ray binaries to confirm or rule out the assumption of dark-matter spike surrounding the stellar-mass BHs. The possible detection of GW signal modified by a dark-matter spike could also help us to discriminate different theories of gravity and constrain the nature of dark-matter particle \citep{hann20}.
\subsection{Evolution of the spike index}
In our simulations, we take a constant spike index. Similar to supermassive BHs, the donor star in BH X-ray binaries could also
interact with the dark matter via gravitational scattering, which tends to drive dark matter particles
into the BH, reducing the spike index by kinetic heating \citep{gned04,merr04}. The kinetic heating effect would cause the
spike index to evolve to $\gamma=1.5$ if gravitational scattering of stars plays an important role \citep{gned04,merr04}.
This change of the spike index is related
to the heating timescale, which can be expressed as \citep{merr04,chan23}
\begin{equation}
\begin{aligned}
t_{\rm heat}=3.4\times10^{8}\times({\rm ln}\Lambda)^{-1}\left(\frac{M_{\rm BH}}{10~M_{\odot}}\right)^{1/2} \\
\left(\frac{r_{\rm in}}{8.2~\rm pc}\right)^{3/2}\left(\frac{M_{\rm d}}{1~M_{\odot}}\right)^{-1}~\rm yr.
\end{aligned}
\end{equation}

During the evolution of BH LMXBs, the BH masses and $r_{\rm in}$ should increase because of an accretion, resulting in
an increasing heating timescale. Meanwhile, the decreasing of the donor-star mass would also result in an increasing
heating timescale. In figure 6, we plot a comparison between the heating timescale and the stellar age for BH binaries with $M_{\rm BH,i} = 10~M_{\odot}$, $M_{\rm d,i} = 3~M_{\odot}$, $P_{\rm orb,i} = 3.0$ days, and different spike indices $\gamma$. For simplicity, we take $M_{\rm BH}=10~M_{\odot}$ and $r_{\rm in}=8.2~\rm pc$ to obtain a uniform evolutionary law of heating timescale.
For $\gamma\geq1.7$, the stellar ages are much shorter than the heating timescale, thus, the change of spike index due to
the kinetic heating effect can be ignored. As $\gamma=1.58$, the kinetic heating effect would drive the spike index evolve to
1.5. To evolve toward BH LMXBs, our models require a spike index in the range of $1.7-2.1$. Therefore, the influence of an evolving spike index on the evolution of our simulated BH LMXBs is relatively insignificant.

\subsection{Range of the spike indices}
If the initial dark-matter halo of a galaxy (before the birth of the supermassive BH) appears as an
Navarro, Frenk, and White profile \citep{nava97}, the spike index after the adiabatic growth of the
supermassive BH is $\gamma=7/3$ \citep{quin95,ulli01}. However, the final spike index is $\gamma=1.5$
if the dark-matter density in the initial halo obeys a uniform distribution \citep{quin95,ulli01}. Our calculations
show that there exists a critical spike index ($\gamma=1.58$) over which BH binaries could evolve into BH LMXBs.
If the dynamical friction of dark matter dominates the evolution of BH X-ray binaries, the existence of those systems with long orbital periods implies that a low spike index is possible.
To match the detected effective temperatures of the donor stars in twelve BH LMXBs, the plausible spike indices
are in the range of 1.7 to 2.1. Therefore, the required range of the spike indices that can match the observations
of BH LMXBs is compatible with the theoretical prediction of dark matter.

\begin{figure}
\centering
\includegraphics[width=1.15\linewidth,trim={0 0 0 0},clip]{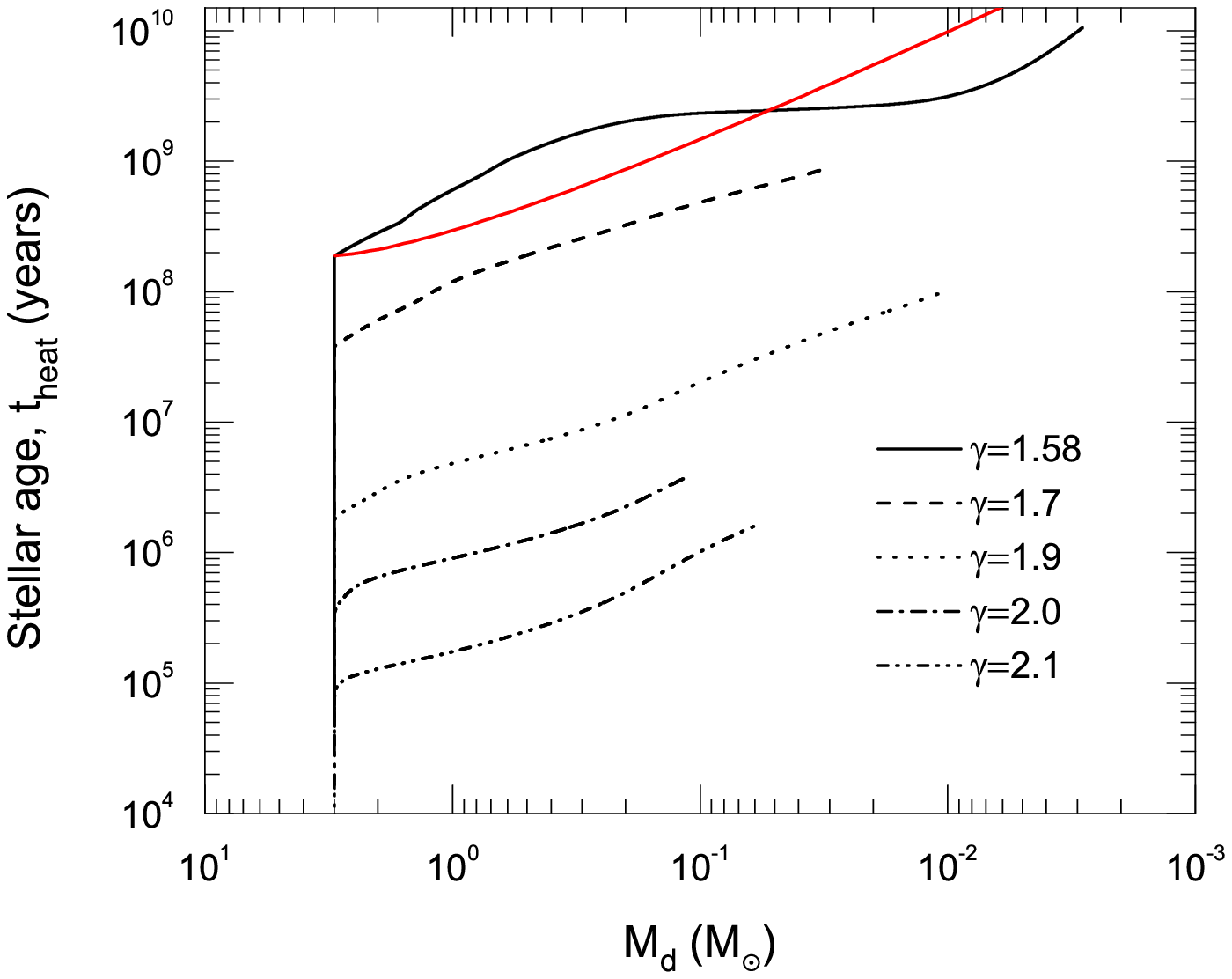}
\caption{Stellar age and heating timescale as a function of the donor-star mass for BH binaries with $M_{\rm BH,i} = 10~M_{\odot}$, $M_{\rm d,i} = 3~M_{\odot}$, $P_{\rm orb,i} = 3.0$ days, and different spike indices $\gamma$. The red and black curves denote the evolutionary track of the heating timescale and stellar age, respectively. } \label{fig:orbmass}
\end{figure}

\section{Conclusion}
In this work, we employ a detailed stellar evolution model to diagnose whether the dynamical friction between dark matter and the companion star can drive BH binaries to evolve toward the observed BH LMXBs and alleviate the effective temperature problem. Our main conclusions are as follows:
\begin{enumerate}
\item If dark matter spikes are present around stellar-mass BHs with the appropriate density distributions, then they would play an important role in dominating the orbital evolution of BH X-ray binaries. Compared with the changes in dark-matter density, the rate of orbital AML due to the dynamical friction is more sensitive to the mass change of the donor star.
\item A high spike index $\gamma$ produces a high rate of orbital AML, resulting in a high mass-transfer rate and a short lifetime of BH X-ray binaries. There exists a critical spike index $\gamma=1.58$, over which the dynamical friction could drive BH binaries to evolve toward BH LMXBs.
\item Considering the dynamical friction, the initial BH masses, initial donor-star masses, and initial orbital periods also influence the evolution of BH X-ray binaries. For a small $\gamma$, BH binaries with a high donor-star mass ($\ga4.0~M_{\odot}$) and a long orbital period are difficult to evolve toward BH LMXBs, while they could for a large $\gamma$.

\item For a high spike index of $\gamma = 2.0$ or $2.1$, our simulated mass-transfer rates are too high to match the observed properties that BH LMXBs generally appears as soft X-ray transients. Therefore, it is most unlikely that the dynamical friction of dark matter could be responsible for the formation of those BH LMXBs with a long orbital period ($>0.3~\rm days$).

\item The dynamical friction can only alleviate the effective temperature problem of the donor stars in those BH LMXBs with a short orbital period ($<0.3~\rm days$). When $\gamma = 1.7-1.9$, our simulated effective temperatures of the donor stars are in good agreement with the observations of some BH LMXBs. A large $\gamma$ tends to form BH LMXBs with long orbital periods, while a small $\gamma$ will produce those BH LMXBs with short orbital periods and low donor-star effective temperatures. Furthermore, our detailed stellar evolution models indicate that the required spike indices of XTE J1118+480 and A0620-00 are not consistent with the prediction given by \cite{chan23}.

\item The dark-matter spike surrounding some stellar-mass BHs may be indirectly confirmed or ruled out by detecting the waveform of low-frequency gravitational waves from some BH ultracompact X-ray binaries if the spike index $\gamma\approx1.7$.
\end{enumerate}

\acknowledgments {We are extremely grateful to the anonymous referee for
helpful comments that improved this manuscript. This work was partly supported by the National Natural Science Foundation of China (under grant Nos. 12273014) and the Natural Science Foundation (under grant No. ZR2021MA013) of Shandong Province.}

\end{document}